\pdfoutput=1
\documentclass[12pt,a4paper]{article}

\usepackage{ifthen} 
\usepackage{overpic}
\usepackage{siunitx}
\newboolean{pdflatex}
\setboolean{pdflatex}{true} 
\usepackage{nicefrac}
\newboolean{articletitles}
\setboolean{articletitles}{true} 

\newboolean{uprightparticles}
\setboolean{uprightparticles}{false} 

\def\paperauthors{LHCb collaboration} 
\def\paperasciititle{Observation of the decay chi_1(372) -> J/psi mu+ mu-} 
\def\papertitle{Observation of the decay $\chicone(3872) \rightarrow \jpsi \mumu$} 
\def\paperkeywords{{High Energy Physics}, {LHCb}} 
\def\papercopyright{\the\year\ CERN for the benefit of the LHCb collaboration} 
\def\paperlicence{CC BY 4.0 licence}
\def\paperlicenceurl{https://creativecommons.org/licenses/by/4.0/}

\newcommand{\phd}{\phantom{.}}
\newif\ifEnableSectionTOCLinks
\EnableSectionTOCLinksfalse 

\usepackage[top=1in, bottom=1.25in, left=1in, right=1in]{geometry}

\usepackage{microtype}
\usepackage{lineno}  
\usepackage{xspace} 
\usepackage{caption} 

\usepackage{graphicx}  
\usepackage{color}
\usepackage{colortbl}
\graphicspath{{./figs/}} 

\usepackage{amsmath} 
\usepackage{amssymb}
\usepackage{amsfonts}
\usepackage{upgreek} 

\newcommand*\patchAmsMathEnvironmentForLineno[1]{%
\expandafter\let\csname old#1\expandafter\endcsname\csname #1\endcsname
\expandafter\let\csname oldend#1\expandafter\endcsname\csname
end#1\endcsname
 \renewenvironment{#1}%
   {\linenomath\csname old#1\endcsname}%
   {\csname oldend#1\endcsname\endlinenomath}%
}
\newcommand*\patchBothAmsMathEnvironmentsForLineno[1]{%
  \patchAmsMathEnvironmentForLineno{#1}%
  \patchAmsMathEnvironmentForLineno{#1*}%
}
\AtBeginDocument{%
\patchBothAmsMathEnvironmentsForLineno{equation}%
\patchBothAmsMathEnvironmentsForLineno{align}%
\patchBothAmsMathEnvironmentsForLineno{flalign}%
\patchBothAmsMathEnvironmentsForLineno{alignat}%
\patchBothAmsMathEnvironmentsForLineno{gather}%
\patchBothAmsMathEnvironmentsForLineno{multline}%
\patchBothAmsMathEnvironmentsForLineno{eqnarray}%
}

\usepackage[pdftex,
            pdfauthor={\paperauthors},
            pdftitle={\paperasciititle},
            pdfkeywords={\paperkeywords}]{hyperref}
\usepackage{hyperxmp}
\hypersetup{
    pdfcopyright={Copyright (C) \papercopyright},
    pdflicenseurl={\paperlicenceurl}
}

\usepackage[colorinlistoftodos,textsize=scriptsize]{todonotes}

\usepackage[bottom,flushmargin,hang,multiple]{footmisc}

\usepackage[all]{hypcap} 

\usepackage{xspace}
\usepackage{upgreek}

\def\lhcb   {\mbox{LHCb}\xspace}

\def\bfactories {\mbox{\B Factories}\xspace}

\def\MagUp {\mbox{\em Mag\kern -0.05em Up}\xspace}

\ifthenelse{\boolean{uprightparticles}}%
{

 \def\Pmu         {\ensuremath{\upmu}\xspace}

 \def\Ppi         {\ensuremath{\uppi}\xspace}
 
 \def\Prho        {\ensuremath{\uprho}\xspace}

 \def\Pchi        {\ensuremath{\upchi}\xspace}
 \def\Ppsi        {\ensuremath{\uppsi}\xspace}
 \def\Pomega      {\ensuremath{\upomega}\xspace}

 \def\PDelta      {\ensuremath{\Delta}\xspace}
 \def\PXi         {\ensuremath{\Xi}\xspace}
 \def\PLambda     {\ensuremath{\Lambda}\xspace}
 \def\PSigma      {\ensuremath{\Sigma}\xspace}
 \def\POmega      {\ensuremath{\Omega}\xspace}
 \def\PUpsilon    {\ensuremath{\Upsilon}\xspace}
 \let\oldPi\Pi
 \def\PPi         {\ensuremath{\oldPi}\xspace}

 \def\PB      {\ensuremath{\mathrm{B}}\xspace}
 \def\PD      {\ensuremath{\mathrm{D}}\xspace}
 \def\PJ      {\ensuremath{\mathrm{J}}\xspace}
 \def\PK      {\ensuremath{\mathrm{K}}\xspace}
 \def\Pb      {\ensuremath{\mathrm{b}}\xspace}
 \def\Pc      {\ensuremath{\mathrm{c}}\xspace}

 \def\Ps      {\ensuremath{\mathrm{s}}\xspace}

 \def\thebaroffset{0.0em}
}
{

 \def\Pmu         {\ensuremath{\mu}\xspace}

 \def\Ppi         {\ensuremath{\pi}\xspace}
 
 \def\Prho        {\ensuremath{\rho}\xspace}

 \def\Pchi        {\ensuremath{\chi}\xspace}
 \def\Ppsi        {\ensuremath{\psi}\xspace}
 \def\Pomega      {\ensuremath{\omega}\xspace}
 \mathchardef\PDelta="7101
 \mathchardef\PXi="7104
 \mathchardef\PLambda="7103
 \mathchardef\PSigma="7106
 \mathchardef\POmega="710A
 \mathchardef\PUpsilon="7107
 \mathchardef\PPi="7105
 \def\PB      {\ensuremath{B}\xspace}
 \def\PD      {\ensuremath{D}\xspace}
 \def\PJ      {\ensuremath{J}\xspace}
 \def\PK      {\ensuremath{K}\xspace}
 \def\Pb      {\ensuremath{b}\xspace}
 \def\Pc      {\ensuremath{c}\xspace}

 \def\Ps      {\ensuremath{s}\xspace}

 \def\thebaroffset{0.18em}
}
\newcommand{\offsetoverline}[2][\thebaroffset]{\kern #1\overline{\kern -#1 #2}}%

\makeatletter
\ifcase \@ptsize \relax
  \newcommand{\miniscule}{\@setfontsize\miniscule{4}{5}}
\or
  \newcommand{\miniscule}{\@setfontsize\miniscule{5}{6}}
\or
  \newcommand{\miniscule}{\@setfontsize\miniscule{5}{6}}
\fi
\makeatother

\DeclareRobustCommand{\optbar}[1]{\shortstack{{\miniscule (\rule[.5ex]{1.25em}{.18mm})}
  \\ [-.7ex] $#1$}}

\def\mup        {{\ensuremath{\Pmu^+}}\xspace}
\def\mun        {{\ensuremath{\Pmu^-}}\xspace} 
\def\mupm       {{\ensuremath{\Pmu^\pm}}\xspace}

\def\mumu       {{\ensuremath{\Pmu^+\Pmu^-}}\xspace}

\def\squark    {{\ensuremath{\Ps}}\xspace}

\def\cquark    {{\ensuremath{\Pc}}\xspace}

\def\bquark    {{\ensuremath{\Pb}}\xspace}

\def\pion   {{\ensuremath{\Ppi}}\xspace}

\def\pip    {{\ensuremath{\pion^+}}\xspace}
\def\pim    {{\ensuremath{\pion^-}}\xspace}
\def\pipm   {{\ensuremath{\pion^\pm}}\xspace}

\def\rhomeson {{\ensuremath{\Prho}}\xspace}
\def\rhoz     {{\ensuremath{\rhomeson^0}}\xspace}

\def\kaon    {{\ensuremath{\PK}}\xspace}

\def\KorKbar {\kern \thebaroffset\optbar{\kern -\thebaroffset \PK}{}\xspace}

\def\Kp      {{\ensuremath{\kaon^+}}\xspace}

\def\D       {{\ensuremath{\PD}}\xspace}

\def\DorDbar {\kern \thebaroffset\optbar{\kern -\thebaroffset \PD}\xspace}

\def\Dp      {{\ensuremath{\D^+}}\xspace}
\def\Dm      {{\ensuremath{\D^-}}\xspace}

\def\DpDm    {\ensuremath{\Dp {\kern -0.16em \Dm}}\xspace}

\def\B       {{\ensuremath{\PB}}\xspace}

\def\BorBbar {\kern \thebaroffset\optbar{\kern -\thebaroffset \PB}\xspace}

\def\Bd      {{\ensuremath{\B^0}}\xspace}

\def\BdorBdbar {\kern \thebaroffset\optbar{\kern -\thebaroffset \Bd}\xspace}
\def\Bu      {{\ensuremath{\B^+}}\xspace}

\def\Bs      {{\ensuremath{\B^0_\squark}}\xspace}

\def\BsorBsbar {\kern \thebaroffset\optbar{\kern -\thebaroffset \Bs}\xspace}

\def\jpsi     {{\ensuremath{{\PJ\mskip -3mu/\mskip -2mu\Ppsi}}}\xspace}
\def\psitwos  {{\ensuremath{\Ppsi{(2S)}}}\xspace}

\def\chicone  {{\ensuremath{\Pchi_{\cquark 1}}}\xspace}
\def\chictwo  {{\ensuremath{\Pchi_{\cquark 2}}}\xspace}

\def\Y#1S{\ensuremath{\PUpsilon{(#1S)}}\xspace}

\def\theX     {{\ensuremath{\Pchi_{c1}(3872)}}\xspace}

\def\LorLbar     {\kern \thebaroffset\optbar{\kern -\thebaroffset \PLambda}\xspace}

\def\BF         {{\ensuremath{\mathcal{B}}}\xspace}

\newcommand{\decay}[2]{\mbox{\ensuremath{#1\!\to #2}}\xspace}

\def\to                 {\ensuremath{\rightarrow}\xspace}

\def\AT#1     {\ensuremath{A_{\mathrm{T}}^{#1}}\xspace}           

\def\C#1      {\ensuremath{\mathcal{C}_{#1}}\xspace}                       
\def\Cp#1     {\ensuremath{\mathcal{C}_{#1}^{'}}\xspace}                    
\def\Ceff#1   {\ensuremath{\mathcal{C}_{#1}^{\mathrm{(eff)}}}\xspace}        
\def\Cpeff#1  {\ensuremath{\mathcal{C}_{#1}^{'\mathrm{(eff)}}}\xspace}       
\def\Ope#1    {\ensuremath{\mathcal{O}_{#1}}\xspace}                       
\def\Opep#1   {\ensuremath{\mathcal{O}_{#1}^{'}}\xspace}                    

\newcommand{\aunit}[1]{\ensuremath{\text{\,#1}}}

\newcommand{\tev}{\aunit{Te\kern -0.1em V}\xspace}
\newcommand{\gev}{\aunit{Ge\kern -0.1em V}\xspace}
\newcommand{\mev}{\aunit{Me\kern -0.1em V}\xspace}
\newcommand{\kev}{\aunit{ke\kern -0.1em V}\xspace}
\newcommand{\ev}{\aunit{e\kern -0.1em V}\xspace}

\newcommand{\mevc}{\ensuremath{\aunit{Me\kern -0.1em V\!/}c}\xspace}
\newcommand{\gevc}{\ensuremath{\aunit{Ge\kern -0.1em V\!/}c}\xspace}
\newcommand{\mevcc}{\ensuremath{\aunit{Me\kern -0.1em V\!/}c^2}\xspace}
\newcommand{\gevcc}{\ensuremath{\aunit{Ge\kern -0.1em V\!/}c^2}\xspace}

\def\fb   {\ensuremath{\aunit{fb}}\xspace}
\def\invfb   {\ensuremath{\fb^{-1}}\xspace}

\def\gsim{{~\raise.15em\hbox{$>$}\kern-.85em
          \lower.35em\hbox{$\sim$}~}\xspace}
\def\lsim{{~\raise.15em\hbox{$<$}\kern-.85em
          \lower.35em\hbox{$\sim$}~}\xspace}

\def\evtgen     {\mbox{\textsc{EvtGen}}\xspace}

\def\geant      {\mbox{\textsc{Geant4}}\xspace}

\def\pythia     {\mbox{\textsc{Pythia}}\xspace}

\def\tell1  {TELL1\xspace}
\def\ukl1   {UKL1\xspace}

\newcommand{\ie}{\mbox{\itshape i.e.}\xspace}

\newcommand{\phz}{\phantom{0}}
\newcommand{\lhcborcid}[1]{\href{https://orcid.org/#1}{\hspace*{0.1em}\raisebox{-0.45ex}{\includegraphics[width=1em]{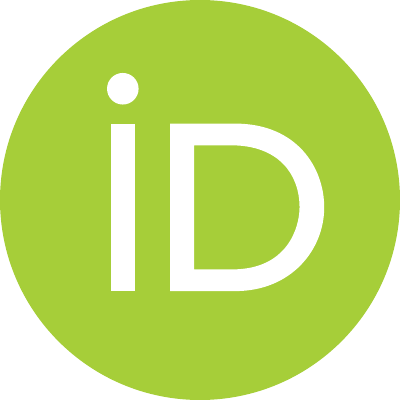}}}}

\hypersetup{
  colorlinks   = true, 
  urlcolor     = blue, 
  linkcolor    = blue, 
  citecolor    = red   
}

\ifEnableSectionTOCLinks
    \usepackage[explicit]{titlesec} 
    
    \let\oldcontentsline\contentsline
    \renewcommand

    \titleformat{\section}{\normalfont\Large\bf}{\hyperlink{tocsection.\thesection}{{\thesection} \parbox[t]{\dimexpr\textwidth-1pc}{#1}}}{1pc}{}

    \titleformat{\subsection}{\normalfont\bf}{\hyperlink{tocsubsection.\thesubsection}{{\thesubsection} \parbox[t]{\dimexpr\textwidth-1pc}{#1}}}{1pc}{}

    \titleformat{name=\section,numberless}[display]{}{}{0pt}{\normalfont\Huge\bfseries #1}
\fi

\usepackage{cite} 
\usepackage{mciteplus}

\usepackage{longtable} 
\usepackage{multirow}

\begin{document}

\renewcommand{\thefootnote}{\fnsymbol{footnote}}
\setcounter{footnote}{1}


\begin{titlepage}
\pagenumbering{roman}

\vspace*{-1.5cm}
\centerline{\large EUROPEAN ORGANIZATION FOR NUCLEAR RESEARCH (CERN)}
\vspace*{1.5cm}
\noindent
\begin{tabular*}{\linewidth}{lc@{\extracolsep{\fill}}r@{\extracolsep{0pt}}}
\ifthenelse{\boolean{pdflatex}}
{\vspace*{-1.5cm}\mbox{\!\!\!\includegraphics[width=.14\textwidth]{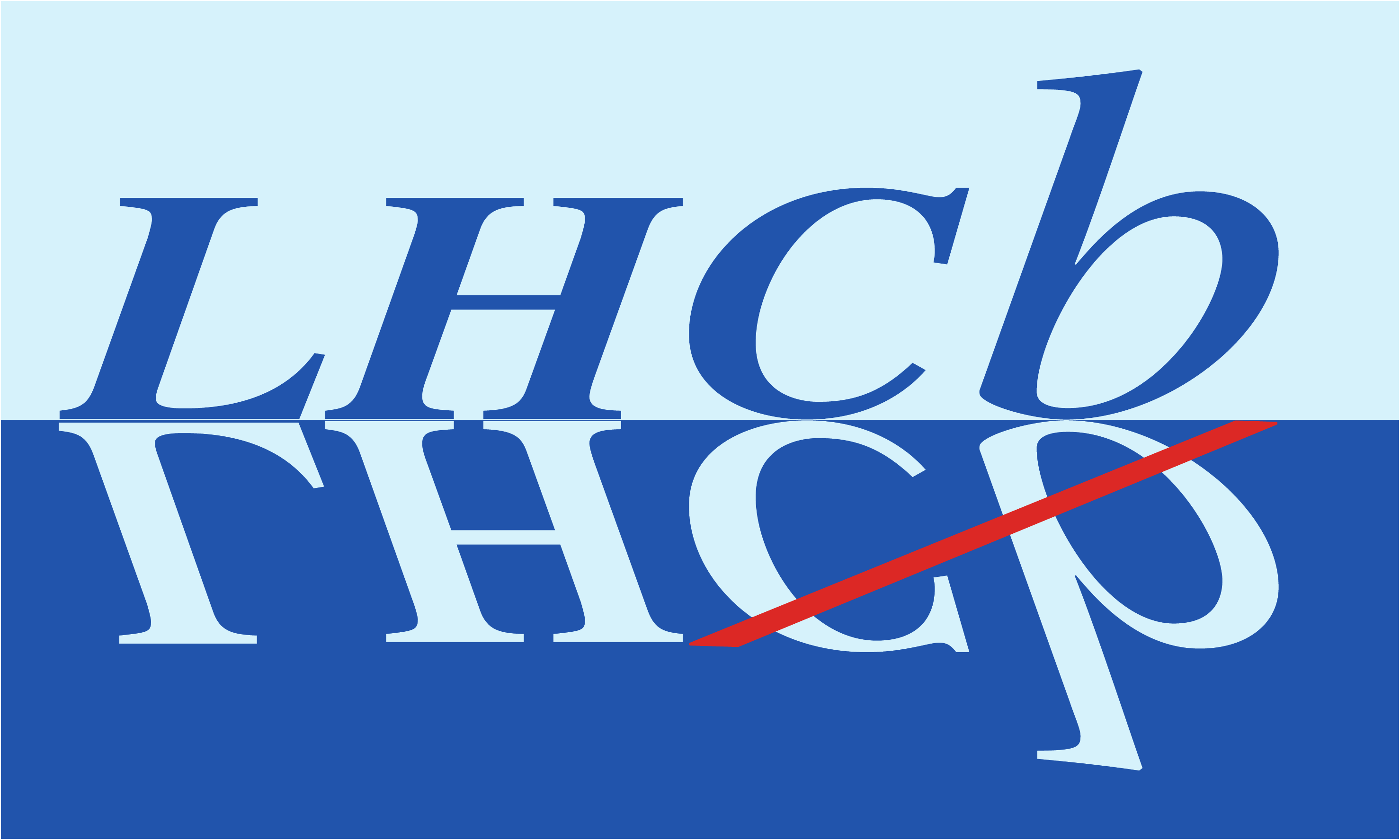}} & &}%
{\vspace*{-1.2cm}\mbox{\!\!\!\includegraphics[width=.12\textwidth]{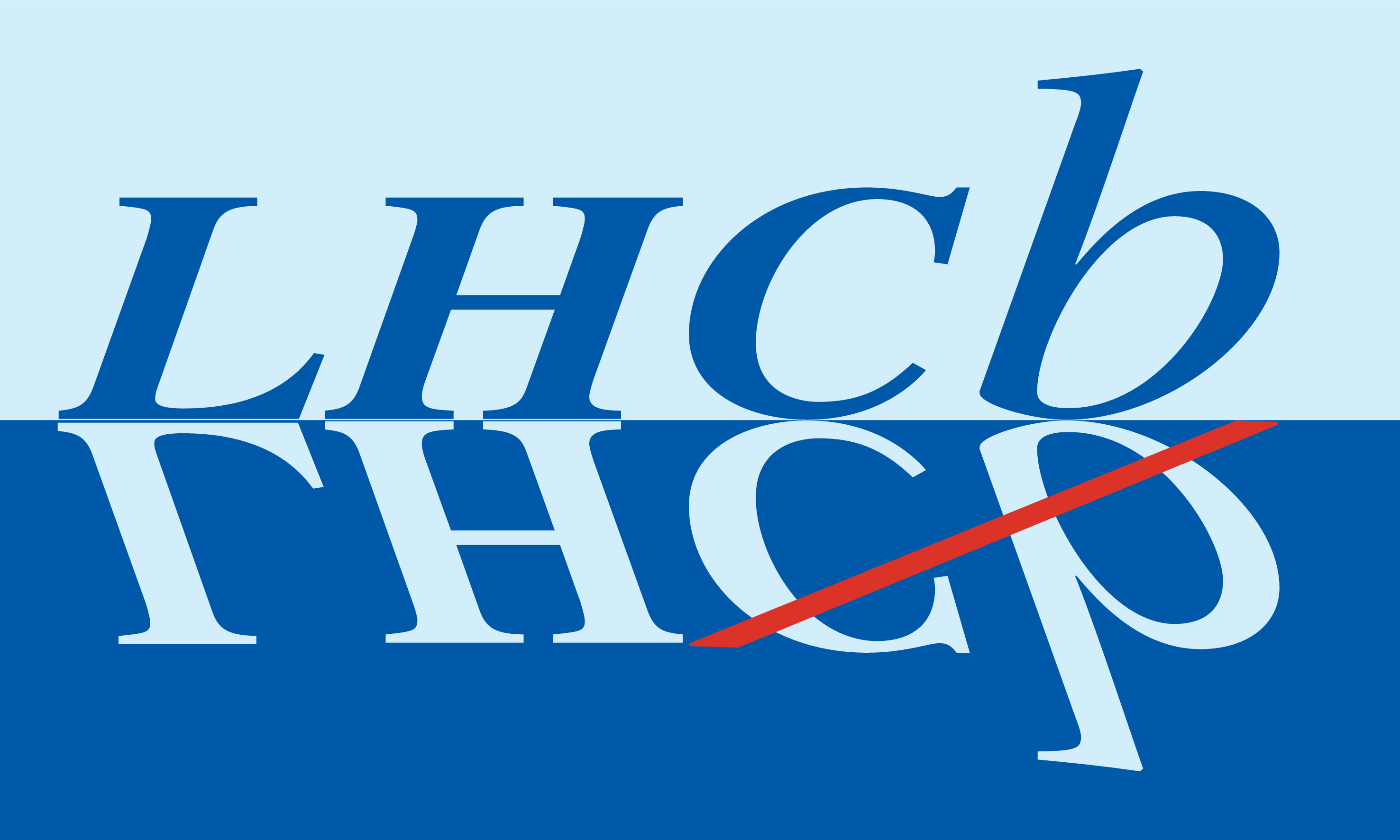}} & &}%
\\
 & & CERN-EP-2025-271 \\  
 & & LHCb-PAPER-2025-043 \\  
 & & May 5, 2026 \\ 
 & & \\
\end{tabular*}

\vspace*{4.0cm}

{\normalfont\bfseries\boldmath\huge
\begin{center}
  \papertitle 
\end{center}
}

\vspace*{2.0cm}

\begin{center}
\paperauthors\footnote{Authors are listed at the end of this paper.}
\end{center}

\vspace{\fill}

\begin{abstract}
  \noindent
    The first observation of the \decay{\theX}{\jpsi\mup\mun} decay is reported using proton-proton collision data recorded with the \lhcb detector corresponding to an integrated luminosity of 9\invfb . The decay mode is observed for the first time, with a significance of $6.5\sigma$. Its branching fraction is measured relative to the \decay{\theX}{\jpsi\pip\pim} decay mode
 \begin{align*}
     \frac{\BF(\decay{\theX}{\jpsi\mup\mun})}{\BF(\decay{\theX}{\jpsi\pip\pim})} = \left(1.68\pm 0.32\pm 0.05\right)\times10^{-3},
\end{align*}
 where the first uncertainty includes both statistical contributions and systematic contributions which are uncorrelated between data-taking periods, and the second represents the systematic contributions that are correlated between data-taking periods. 
\end{abstract}

\vspace*{2.0cm}

\begin{center}
  Published in Phys. Rev. D 113 (2026) 072019
 
\end{center}

\vspace{\fill}

{\footnotesize 
\centerline{\copyright~\papercopyright. \href{\paperlicenceurl}{\paperlicence}.}}
\vspace*{2mm}

\end{titlepage}


\newpage
\setcounter{page}{2}
\mbox{~}
%
%
%
%


\renewcommand{\thefootnote}{\arabic{footnote}}
\setcounter{footnote}{0}


\cleardoublepage


\pagestyle{plain} 
\setcounter{page}{1}
\pagenumbering{arabic}


\section{Introduction}
\label{sec:Introduction}
 
Over the last two decades, there has been growing interest in the spectroscopy of states that do not fit the conventional charmonium spectrum~\cite{husken2025briefguideexotichadrons}. This was sparked by the observation of the charmonium-like $X(3872)$ state, now referred to as $\chicone(3872)$, by the Belle collaboration in 2003~\cite{Choi:2003ue}. 
Although the existence of the $\chicone(3872)$ state has been confirmed by many experiments~\cite{Abazov:2004kp, Aubert:2008gu, Aaltonen:2009vj, LHCb-PAPER-2011-034, Ablikim:2013dyn} and its quantum numbers measured to be $1^{++}$~\cite{LHCb-PAPER-2013-001}, its nature is still disputed. It may be an exotic state, such as a compact tetraquark~\cite{Maiani:2004vq}, loosely bound deuteron-like $D^{*0}{\overline{D}}{}^0$ `molecule'~\cite{Tornqvist:2004qy} or a charmonium-molecule mixture~\cite{Hanhart:2011jz}.  Hence, more studies are needed to elucidate the nature of the \theX state.

The \decay{\theX}{\jpsi\mup\mun} decay has several possible contributions depending on the mass of the dimuon pair, $m_{\mumu}$.  At low $m_{\mumu}$, the virtual photon pole dominates the decay rate, which is directly related to the radiative decay of the $\theX$ state \cite{LHCb-PAPER-2024-015}. At higher dimuon mass, real $\rhoz$ and $\omega$ meson production are expected as has been seen in the decay \decay{\theX}{\jpsi\pip\pim} \cite{LHCb-PAPER-2021-045}. The $\omega$ meson couples more strongly to a dimuon pair than the $\rho^0$ meson \cite{PDG2024}. Consequently, the $\omega$ meson contribution to the final state in the $\jpsi \mup \mun$ mode is potentially more prevalent than in the $\jpsi \pip \pim$ mode. So far, the \decay{\theX}{\jpsi\mup\mun} decay has received little attention in the literature. Currently, the only available study is that of Ref.~\cite{Colangelo:2025yud}, in which the radiative decay width and transition form factor for analogous decay modes are employed to estimate the Dalitz contribution. The resulting branching fraction is predicted to be $(4.2 \pm 1.7) \times 10^{-5}$.
 
 Above the $\theX$ mass other states may be present. In particular, the $\chi_{c0}(3915)$ is a potentially exotic state with a relatively narrow width seen by the \bfactories
 decaying to the $\jpsi \omega$ final state~\cite{Abe:2004zs,Uehara:2009tx,delAmoSanchez:2010jr,Lees:2012xs}. Since it decays to $\jpsi\omega$, this state may be observed in the $\jpsi \mup \mun$ invariant-mass distribution for dimuon masses above the $\omega$ mass threshold. Though the present dataset is not large enough to determine which contributions are present, they need to be considered when modelling background in the sideband region.

 In this paper, the observation of the \decay{\theX}{\jpsi\mup\mun} (with \decay{\jpsi}{\mup\mun}) decay mode is reported for the first time and the branching ratio with respect to the abundant \jpsi\pip\pim mode is measured. Displaced \theX candidates originating from the decay of long-lived beauty hadrons are used, \ie ones that are not coming directly from the proton-proton collision vertex. The measurement is based on data collected by the LHCb collaboration in proton-proton collisions between 2011 and 2018. This dataset corresponds to an integrated luminosity of 3\invfb collected at centre-of-mass energies of $\sqrt{s} = 7$ and $8 \tev$ (Run~1) and 6\invfb collected at $13~\tev$ (Run~2).\footnote{Natural units with $\hbar=c=1$ are used throughout.} This observation builds on the observation of similar muon Dalitz decays of the $\chi_c$ and $\chi_b$ mesons \cite{LHCb-PAPER-2024-025,LHCb-PAPER-2017-036}.
 
\section{Detector and simulation}
\label{sec:Detector}
The \lhcb detector~\cite{LHCb-DP-2012-002,LHCb-DP-2014-002} is a single-arm forward spectrometer covering the \mbox{pseudorapidity} range $2<\eta <5$, designed for the study of particles containing \bquark or \cquark quarks. The high-precision tracking system provides a measurement of the momentum of charged particles with a relative uncertainty that varies from 0.5\% at low momentum to 1.0\% at $200 \gev$. Large samples of  \decay{\jpsi}{\mup\mun} and \decay{\Bu}{\jpsi\Kp} decays, collected concurrently with the dataset used in this analysis, are used to calibrate the momentum scale of the spectrometer~\cite{LHCb-PAPER-2013-011}.\footnote{Throughout this paper, $\rhoz$ and $\omega$ denote the $\rho(770)^0$ and $\omega(782)$ mesons, respectively.} According to these studies, for $\bquark$-hadron decay modes such as the latter, the mass resolution agrees between data and simulation to better than 10\%.

Various charged hadrons are distinguished using information from two ring-imaging Cherenkov (RICH) detectors. In addition, photons, electrons, and hadrons are identified by a calorimeter system consisting of scintillating-pad and preshower detectors, an electromagnetic and a hadronic calorimeter. Muons are identified by a system composed of alternating layers of iron and multiwire proportional chambers.

The online event selection is performed by a trigger, which consists of a hardware stage followed by a two-level software stage~\cite{LHCb-DP-2019-001}. The first stage of the software trigger performs a partial event reconstruction and requires events to have two well-identified oppositely charged muons with an invariant mass larger than $2.7 \gev$ without biasing the decay-time distribution. The second stage performs a 
full event reconstruction.  Events are retained for further 
processing if they contain a displaced $\jpsi \rightarrow \mu^+ \mu^-$
candidate. The  $\jpsi$  decay vertex is required to be well separated 
from each reconstructed primary vertex (PV) of the proton-proton interaction by requiring the distance between the PV and the decay vertex to be significant. 
Triggered data further undergo a centralised, offline processing step
to deliver physics-analysis-ready data across the entire \lhcb physics programme~\cite{Stripping}.

Simulated samples are used in order to model the resolution of the signal and normalisation decays, as well as estimate the efficiency of the selection. In  simulation, proton-proton collisions are generated using
\pythia~\cite{Sjostrand:2007gs}  with a specific
\lhcb configuration~\cite{LHCb-PROC-2010-056}.  Decays of hadronic
particles are described by \evtgen~\cite{Lange:2001uf}. The interaction of the generated
particles with the detector, and its response, are implemented using the \geant toolkit~\cite{Allison:2006ve,
  *Agostinelli:2002hh} as described in Ref.~\cite{LHCb-PROC-2011-006}. 

\section{Selection}
\label{sec:selection}
The selection for both the \decay{\theX}{\jpsi\mup\mun} signal mode and the \decay{\theX}{\jpsi\pip\pim} normalisation mode starts from displaced \jpsi candidates selected by the high-level trigger. In addition to the trigger requirements, the mass of the dimuon pair from the \jpsi decay is required to be within the range $3040 < m_{\mup\mun} < 3150\mev$ and each muon must fulfil tighter requirements on the track quality and muon identification.

\begin{figure}
    \centering
    \begin{overpic}[width=0.98\linewidth]{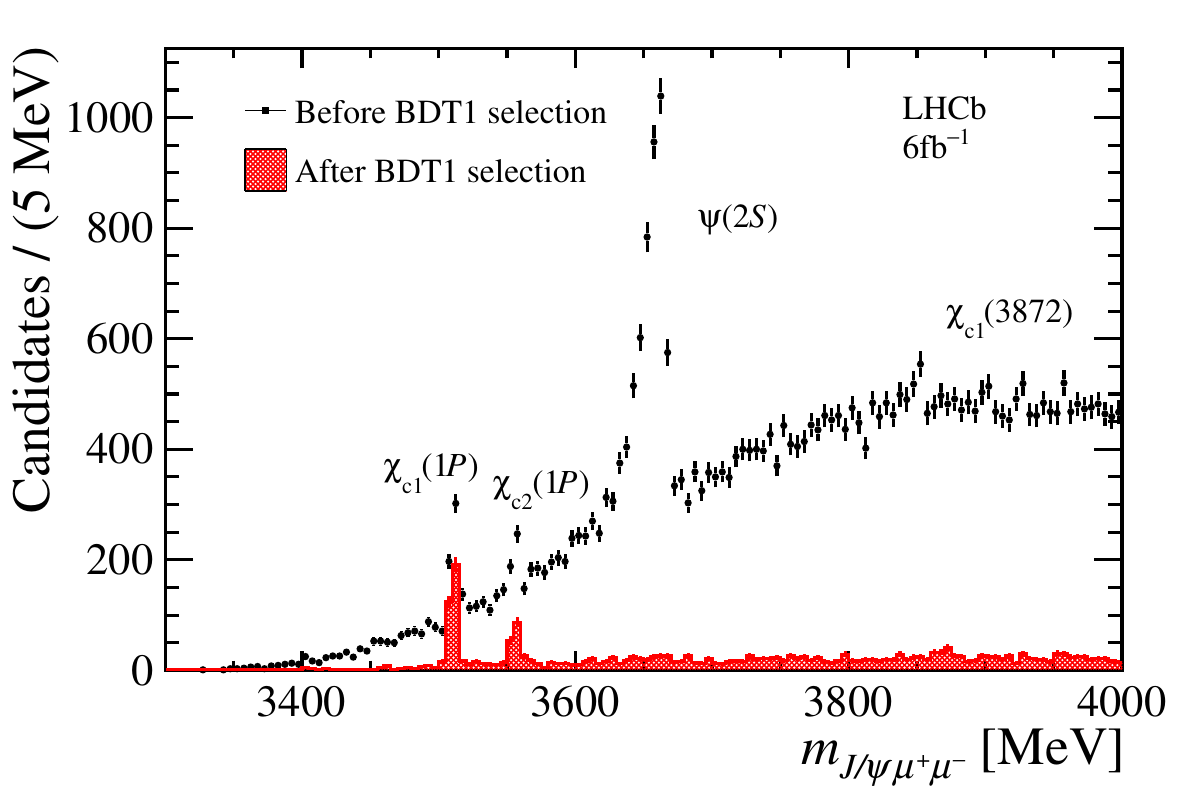}
    \end{overpic}
    \caption{Mass distribution of \jpsi\mup\mun candidates in Run~2 data. From left to right the $\chicone(1P)$, $\chictwo(1P)$, \psitwos ($\rightarrow \jpsi \pip \pim$ with pions misidentified as muons) and \theX peaks are labelled.}
    \label{fig:jpsimumuSpectrum}
\end{figure}

To form \decay{\theX}{\jpsi\mup\mun} candidates, the selected $\jpsi$ mesons are combined with a pair of oppositely charged tracks, identified as muons with loose requirements. Same-sign candidates are also considered in both the signal and normalisation modes for selection optimisation (and for background modelling in the mass fit), corresponding to \jpsi\mupm\mupm and \jpsi\pipm\pipm final-state particles for the signal and normalisation modes, respectively. A kinematic vertex fit is applied to the selected candidates~\cite{Hulsbergen:2005pu} in order to improve the \theX invariant-mass resolution. In the fit, the mass of the dimuon pair from the \jpsi decay is constrained to the known value~\cite{PDG2024} and the four muons are required to originate from a common vertex. 
Requirements are also imposed on the reduced $\chi^2$~of this fit, and the flight distance between the primary and secondary vertices is required to be significant. To further reject background from the decay \decay{\theX}{\jpsi \pip \pim} with pions misidentified as muons, a boosted decision tree~\cite{Breiman,AdaBoost} classifier (referred to as BDT1) from the TMVA toolkit~\cite{Hocker:2007ht,*TMVA4} is trained. This classifier uses \decay{\theX}{\jpsi\mup\mun} and \decay{\theX}{\jpsi\pip\pim} simulation samples as signal and background proxies, respectively. Six variables related to the particle identification (PID) and kinematics of the two muons produced by the \theX decay are considered in the training. The performance of the classifier in data and simulation is validated using the \decay{\chicone(1P)}{\jpsi\mup\mun} decay as a control mode. The working point for the classifier is chosen to maximise the significance of the \decay{\chicone(1P)}{\jpsi\mup\mun} decay in data. This requirement keeps 94\% of the signal in the \decay{\theX}{\jpsi\mup\mun} Run~1 and Run~2 simulation samples while rejecting $98\%$ of the background from pions which are misidentified as muons (judged using the \decay{\psitwos}{\jpsi\pip\pim} peak in data). The effect of this selection is shown in Fig.~\ref{fig:jpsimumuSpectrum} for the Run~2 data sample.

Candidates for the \decay{\theX}{\jpsi\pip\pim} mode are selected similarly to the signal decay apart from the PID and BDT1 requirements. The pions for this decay mode are efficiently selected using information from the RICH detectors. In this decay mode, combinatorial background is reduced with high efficiency by requiring an upper bound on the energy release, $m_{\jpsi\pip\pim}-m_{\jpsi}-m_{\pip\pim} < 200\mev$.

The last step of the selection for both the signal and normalisation is to reduce the background resulting from random combination of \jpsi mesons with muons using a second boosted decision tree algorithm (referred to as BDT2). 
This is trained on the combined simulation samples for the signal and normalisation modes as signal proxy and same-sign candidates ($\jpsi \pi^\pm \pi^\pm$) selected in the data as background proxy. Only the normalisation same-sign candidates are used since the corresponding signal sample is much smaller and its omission has a negligible impact on the performance. As input, BDT2 takes ten variables related to the kinematics, track and vertex quality, as well as the displacement of the \theX candidate from the associated primary vertex, defined as the one which matches best to the \theX flight direction. The classifier shows good agreement between simulation and data for the large $\jpsi \pip \pim$ dataset. When applied to the respective simulation samples the optimal working point, chosen to maximise both the significance of the $\decay{\theX}{\jpsi\mup\mun}$ decay and the ratio of efficiencies between the $\decay{\theX}{\jpsi\mup\mun}$ and $\decay{\theX}{\jpsi\pip\pim}$ decays, selects $88\%$ and $93\%$ of the candidates in the Run~1 and Run~2 signal sample, respectively and $89\%$ and $95\%$ of the candidates in the Run~1 and Run~2 normalisation sample, respectively.  

\section{Mass model}
The \decay{\theX}{\jpsi\mup\mun} yield is determined from an extended unbinned maximum-likelihood fit to the \jpsi\mup\mun mass distribution. A phase-space factor is needed to accurately model the distribution~\cite{LHCb-PAPER-2020-008}, defined as
\begin{equation}
    P_{\jpsi} \equiv \frac{\lambda^{\nicefrac{1}{2}}\left(m_{\jpsi\mup\mun}^2, m_{\jpsi}^2, m_{\mup\mun}^2\right)}{2m_{\jpsi\mup\mun}}
\end{equation}
where $\lambda$ is the K\"all\'en function~\cite{Kallen:1964lxa} and $m_{\jpsi\mup\mun}$, $m_{\jpsi}$ and $m_{\mup\mun}$ are the masses of the \jpsi meson and two muons, the mass of the \jpsi and the mass of the dimuon system, respectively. As an approximation, the mass of the dimuon system inputted into the equation is fixed to $500\mev$, the approximate midpoint of the available phase space. Every component of the final probability density function is multiplied by this phase-space factor. However, it only significantly impacts the high dimuon mass spectrum where the variation of the phase space is largest.

The \theX line shape is described by a Breit--Wigner (BW) function~\cite{LHCb-PAPER-2020-008,LHCb-PAPER-2020-009} convoluted with a modified Gaussian function with power-law tails on both sides of the peak (double-sided Crystal Ball~\cite{Skwarnicki:1986xj}, DCSB) whose parameters are fixed from simulation except for the width of the Gaussian core. For this width, the value found in simulation is scaled by the ratio of widths (measured as $f = 1.02\pm0.04$) found in data and simulation for the \jpsi\pip\pim mode. A Gaussian constraint centred around the resulting value is applied in the likelihood fit, taking into account the uncertainty on the scale factor. The width of the BW function is fixed to the known width of the \theX state~\cite{PDG2024} while the mass is allowed to float.

The combinatorial background is modelled by a first-order polynomial function, which is then multiplied by a phase-space factor as previously mentioned. The shape of the combinatorial background is well described by the same-sign \jpsi\mupm\mupm candidates. Further residual backgrounds resulting from any misidentified \decay{\theX}{\jpsi\pip\pim} decays are modelled using another DSCB function, with parameters fixed from simulation. A BW convolved with a DSCB is also included for the $\chi_{c0}(3915)$ state.

The same-sign sample is considered as an additional category in the fit, allowing the background shape in the control sample to constrain that in the signal sample. The Run~1--2 samples are included in the fit model as separate components. This gives a total of four categories in the simultaneous fit. The parameters of background components are shared between the opposite-sign and same-sign samples.

The normalisation channel is fitted independently of the signal channel. The signal component for this decay is modelled by a relativistic BW convolved with DSCB function whose parameters are fixed from simulation, except for the mass of the BW function and the width of the DSCB function. The combinatorial background is modelled with a first-order polynomial function.

\section{Fit results}
\begin{figure}[tb]
    \centering
    \includegraphics[width=0.45\linewidth]{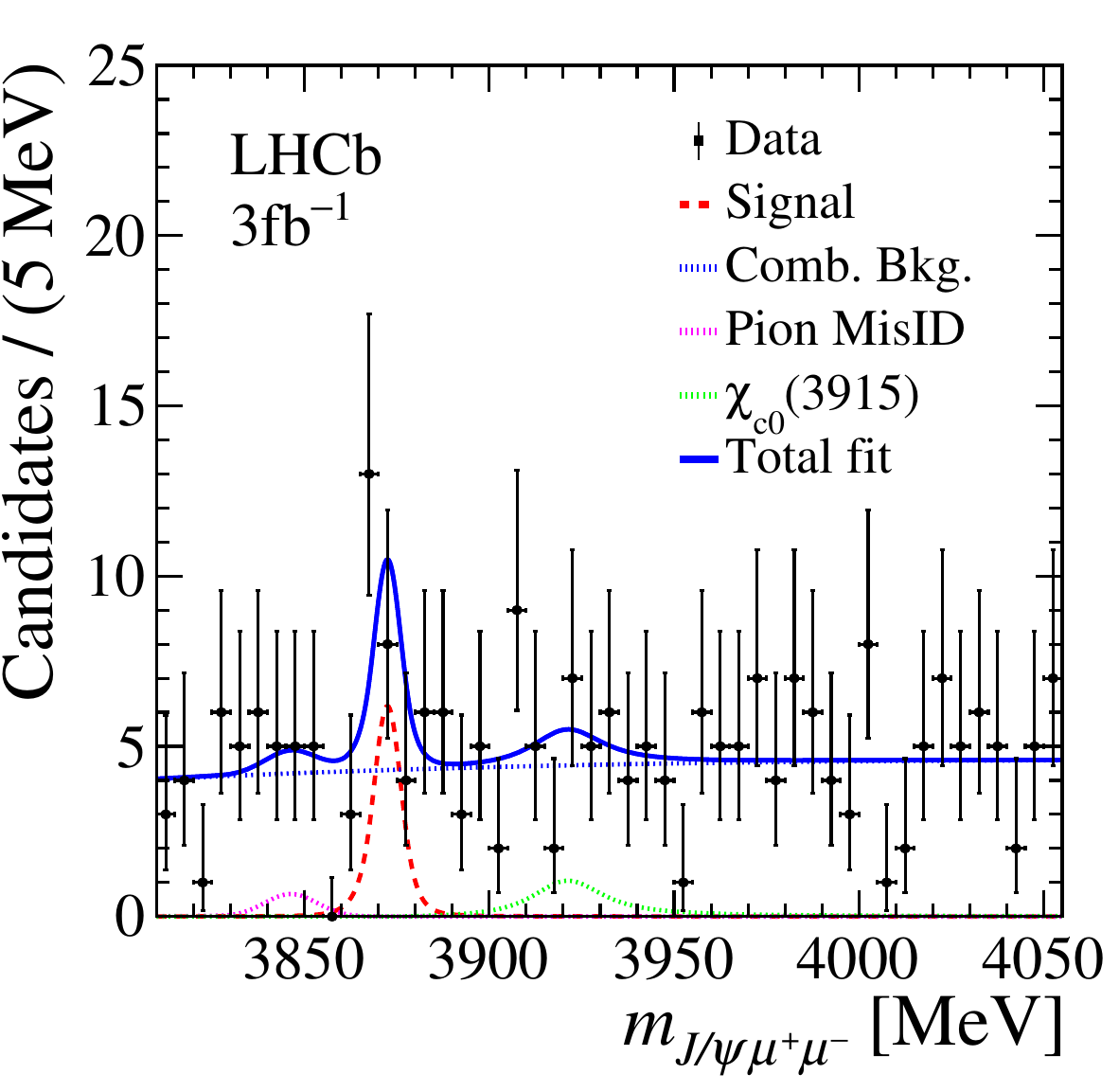}
    \includegraphics[width=0.45\linewidth]{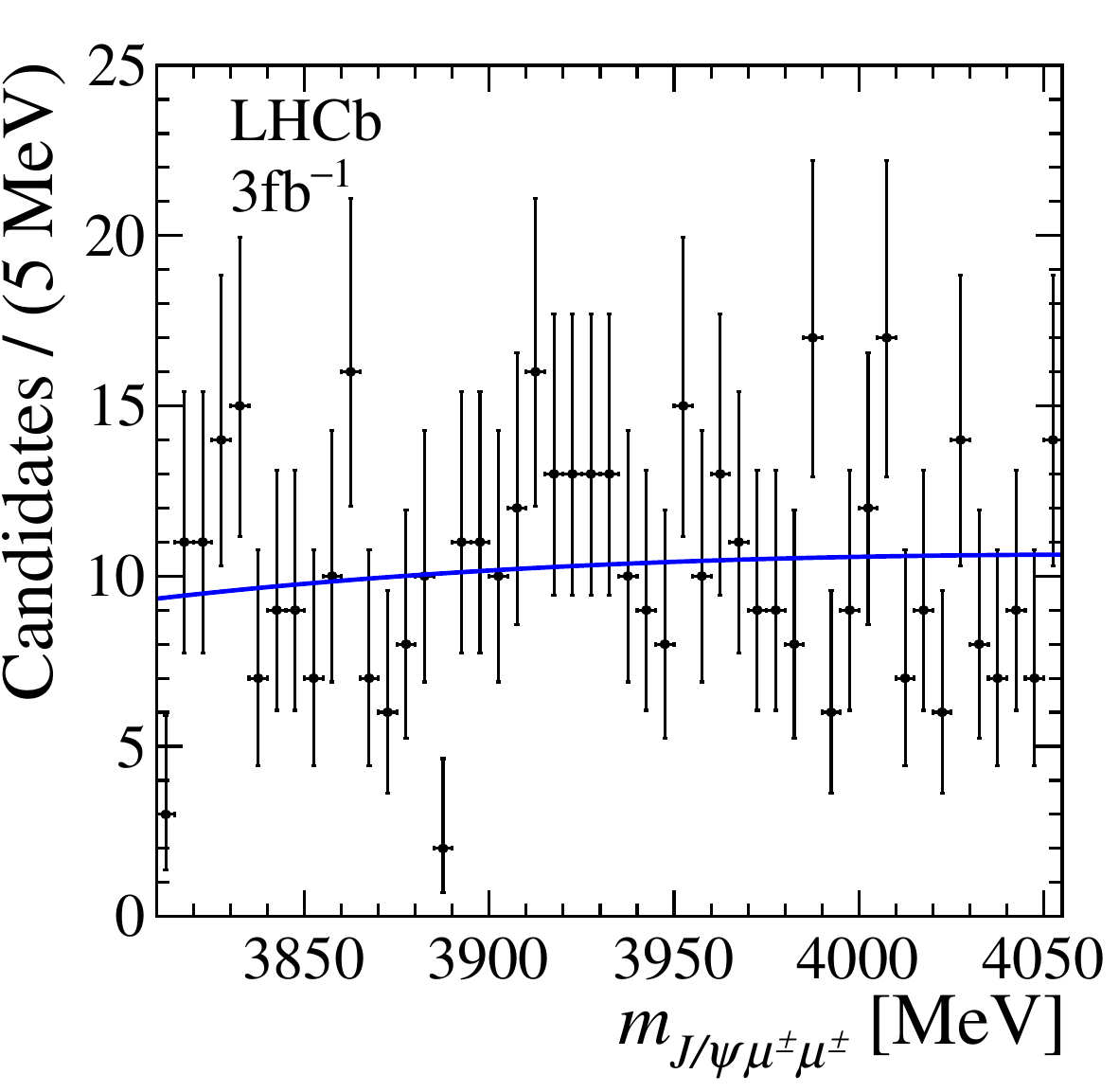}\\
    \includegraphics[width=0.45\linewidth]{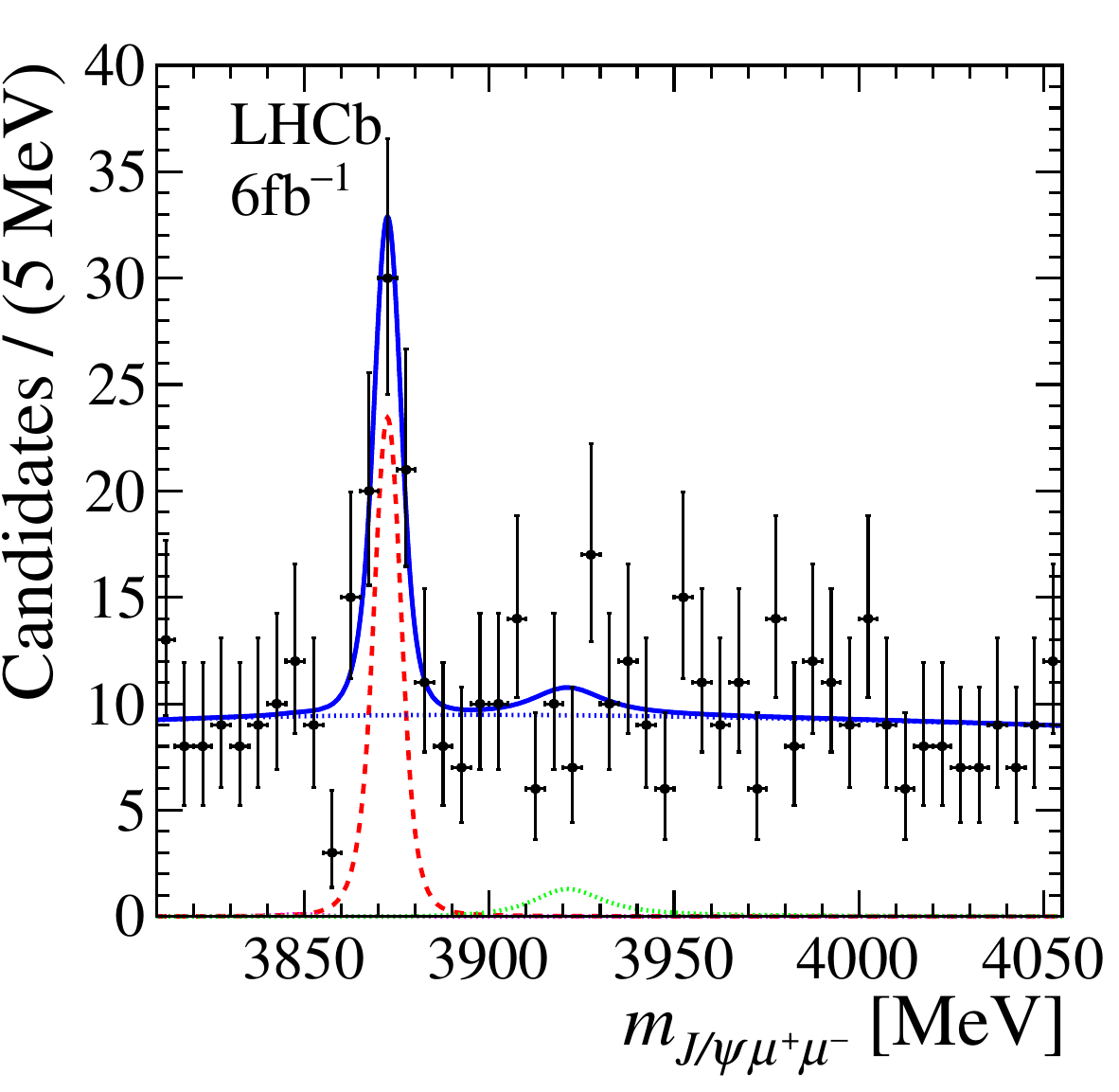}
    \includegraphics[width=0.45\linewidth]{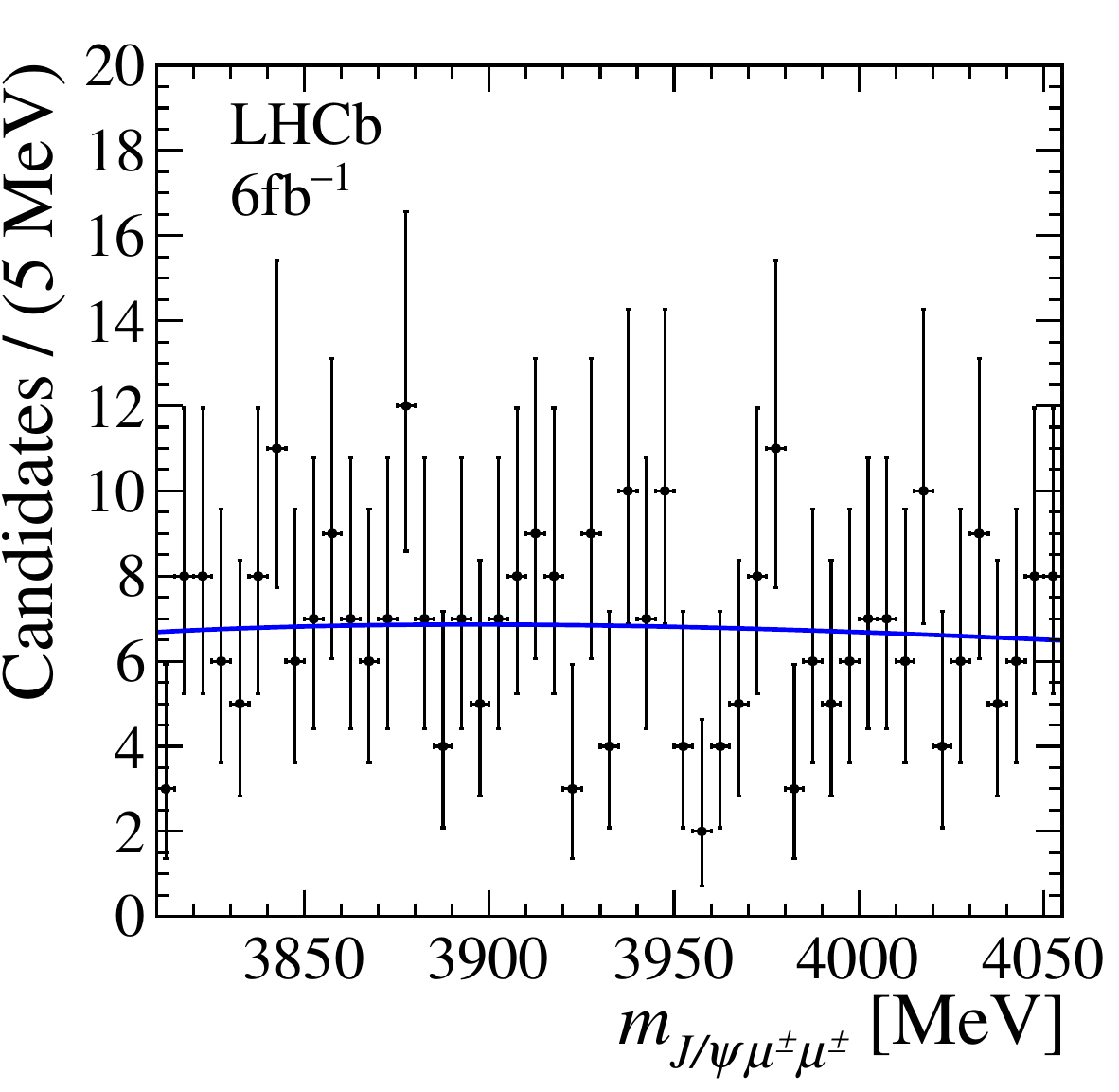}\\
    \caption{Mass distribution for (top) Run~1 and (bottom) Run~2 samples of (left)  $\jpsi\mup\mun$ signal and (right) same-sign $\jpsi\mu^\pm\mu^\pm$ background candidates with the results of the fit also shown.}
    \label{fig:sigFit} 
\end{figure}

\begin{figure}[tb]
    \centering
    \begin{overpic}[width=0.48\linewidth]{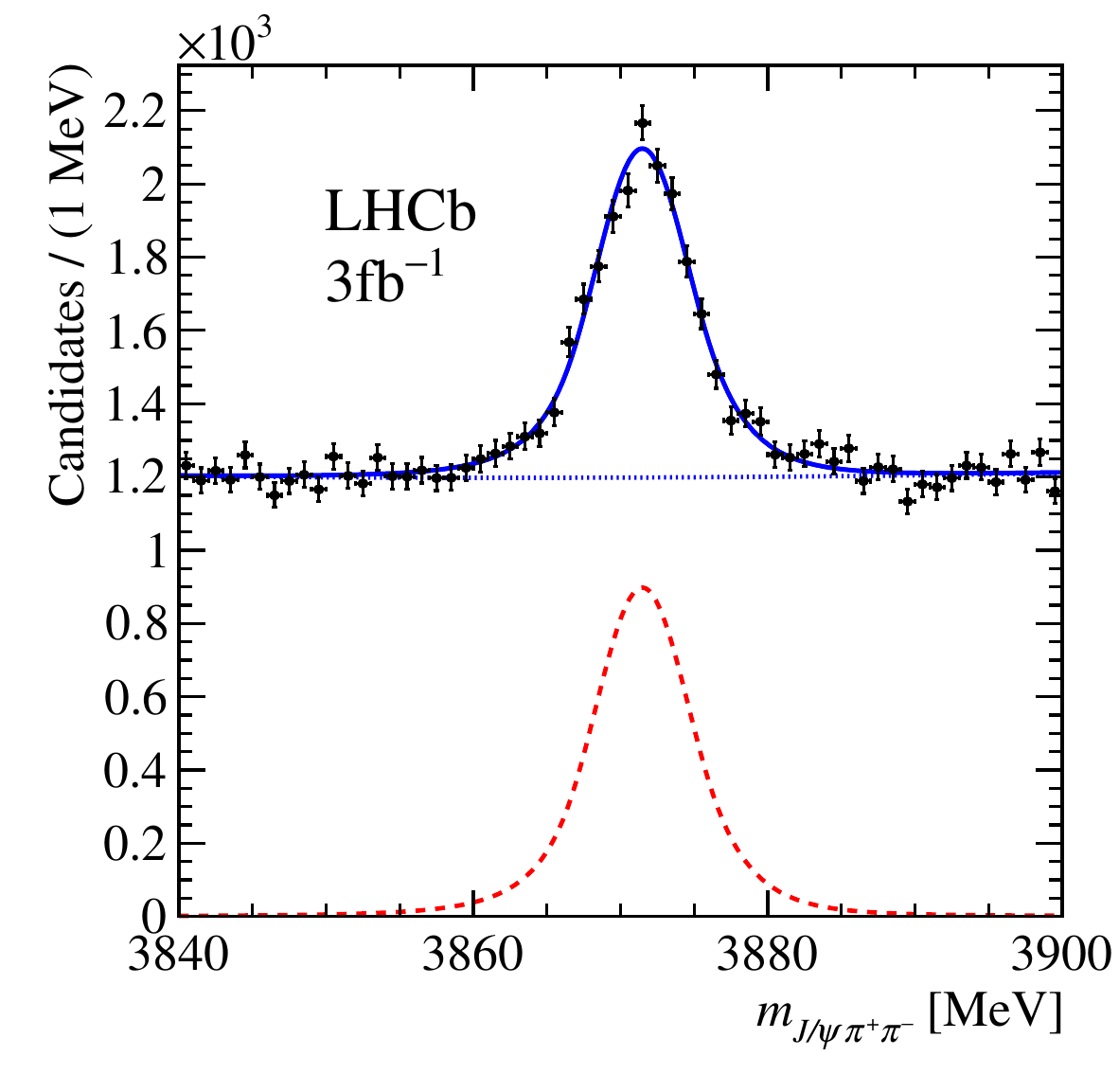}
    \end{overpic}
    \begin{overpic}[width=0.48\linewidth]{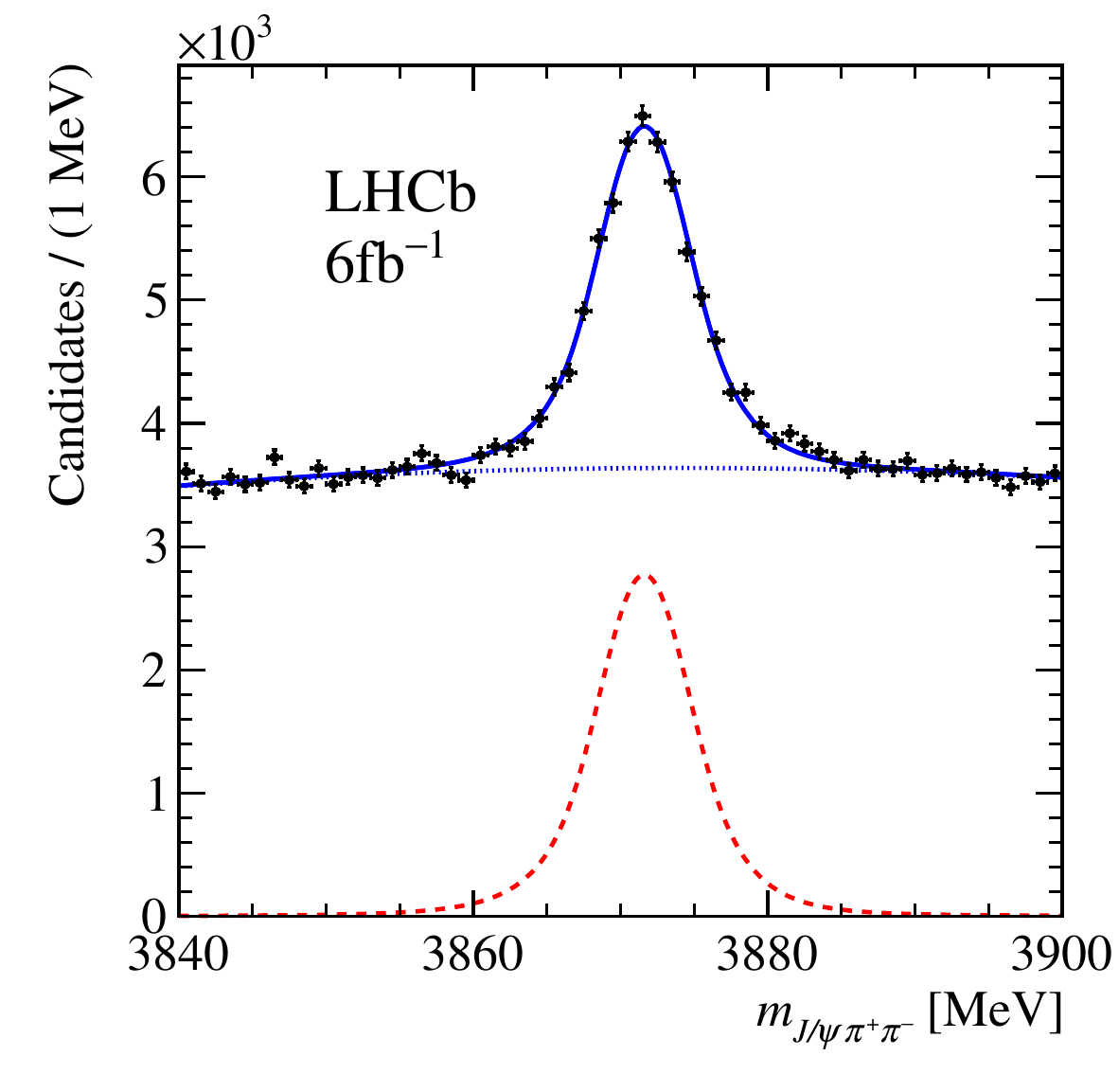}
    \end{overpic}
    \caption{The $m_{\jpsi\pip\pim}$ distribution for (left) Run~1 and (right) Run~2 data with the results of the fit overlaid. The red (blue) dashed line represents the signal (background) component.}
    \label{fig:pipiFit}
\end{figure}

The results of the signal fit are shown in Fig.~\ref{fig:sigFit} with the fit parameters summarised in Table~\ref{tab:sigFit}. The total \decay{\theX}{\jpsi\mup\mun} yield is $60\pm11$ for Run~1--2 cumulatively. The significance of the signal corresponds to $6.5\sigma$ (using Wilks' theorem~\cite{Wilks:1938dza}) including systematic uncertainties due to the mass model (see  Sec.~\ref{sec:systs}), representing the first observation of the \decay{\theX}{\jpsi\mup\mun} decay. The total $\chi_{c0}(3915)$ yield is $14\pm14$, which suggests that the available data does not allow strong conclusions on the presence of the state. A dedicated study of this state could be performed in the future. The results of the normalisation fit are shown in Fig.~\ref{fig:pipiFit}. Here, the only parameter of interest is the total \theX yield, measured to be \SI[separate-uncertainty = true]{35.1(5)e+3}{}.
\begin{table}[tb]
    \centering
    \caption{Results of the simultaneous fit to the full Run~1--2 datasets. Note that `background yield' refers to the background within the signal sample, and thus does not include the same-sign background yield.}
    \label{tab:sigFit}
    \begin{tabular}{cclr@{\:$\pm$\:}}
         Run & Parameter & Fitted Value \\
         \hline
         \multirow{4}{*}{1} & \theX yield & $\phz12.8\pm\phz5.6$\\
          & Combinatorial background yield           & $220\phd\phz\pm20\phz$ \\
          & $\chi_{c0}(3915)$ yield & $\phz\phz6.1\pm\phz7.8$\\
          & Pion misID yield & $\phz\phz2.4\pm5.2$\\
          \hline
          \multirow{4}{*}{2} & \theX yield & $\phz48.6\pm\phz9.7$\\ 
          & Combinatorial background yield           & $460\phd\phz\pm30\phd\phz$\\
          & $\chi_{c0}(3915)$ yield & $\phz\phz8\phd\phz\pm12$\\
          & Pion misID yield & $\phz\phz0\phd\phz\pm61$\\
          \hline
          \multirow{4}{*}{1 and 2} & \theX mass & $3872.58\phz\pm0.83\mev$ \\
          & $\chi_{c0}(3915)$ mass & $3921.8\phz\phz\pm1.8\mev\phz$\\
                                   & \theX scaled mass resolution & $\phz\phz\phz3.45\phz\pm0.13\mev$ \\
                                   
                                   & $\chi_{c0}(3915)$ natural width & $\phz\phz18.9\phz\phz\pm3.5\mev$ \\
    \end{tabular}
\end{table}

\section{Efficiencies}
\label{sec:Effs}
The relative detection and selection efficiencies of the signal and normalisation mode are required to obtain the ratio of branching fractions from the measured event yields. They are evaluated from simulation separately for Run~1 and Run~2. The efficiencies include contributions from detector acceptance, trigger and offline selections. The ratio of efficiencies between signal and normalisation modes is $1.04\pm0.02$ and $1.05\pm0.05$ for Run~1 and Run~2, respectively, where the quoted uncertainty is statistical only.

Corrections to the efficiencies are also defined from each of the BDT-based selections, as well as the PID of the normalisation channel. These are found by comparing the efficiency of the optimal requirement between simulation and data (using the \decay{\chicone}{\jpsi\mup\mun} decay in the case of BDT1 and the \decay{\psitwos}{\jpsi\pip\pim} decay for the normalisation PID) or between the signal and normalisation modes (in the case of BDT2) in simulation. These corrections are summarised in Table~\ref{tab:corrSummary}.
 \begin{table}[tb]
 \caption{Corrections applied to the total efficiency for both Run~1--2, where $\epsilon^{\rm Corr}_{\rm PID}$ represents the correction concerning the PID in the normalisation channel, and $\epsilon^{\rm Corr}_{\rm BDT1}$ and $\epsilon^{\rm Corr}_{\rm BDT2}$ represent the corrections concerning the two BDTs. The uncertainties are statistical only.}
     \label{tab:corrSummary}
     \centering
     \begin{tabular}{ccc}
         \multirow{2}{*}{Correction} & \multicolumn{2}{c}{Value} \\
           & Run~1 & Run~2 \\
           \hline
           $\epsilon^{\rm Corr}_{\rm PID}$  & $0.9896\pm0.0006$ & $1.0072\pm0.0004$ \\
           $\epsilon^{\rm Corr}_{\rm BDT1}$  & $0.96\phz\phz\pm0.03\phz\phz$ & $1.035\phz\pm0.014\phz$ \\
           $\epsilon^{\rm Corr}_{\rm BDT2}$ & $0.9884\pm0.0018$ & $0.9816\pm0.0017$ \\
     \end{tabular}
 \end{table}
 
 The ratio of branching fractions between the signal and normalisation channels is defined as
\begin{equation}
    \mathcal{R} \equiv \frac{\BF(\decay{\theX}{\jpsi\mup\mun})}{\BF(\decay{\theX}{\jpsi\pip\pim})} = \frac{N_{\mumu}\;\epsilon_{\pip\pim}\epsilon^{\rm Corr}_{\rm PID}}{N_{\pip\pim}\;\epsilon_{\mumu}\epsilon^{\rm Corr}_{\rm BDT1}\epsilon^{\rm Corr}_{\rm BDT2}},
    \label{eq:BFeq}
\end{equation}
where $\epsilon_{\mumu} (\epsilon_{\pip\pim})$ is the total efficiency for the signal (normalisation) mode, and $\epsilon^{\rm Corr}_{\rm PID}$, $\epsilon^{\rm Corr}_{\rm BDT1}$ and $\epsilon^{\rm Corr}_{\rm BDT2}$ are the three corrections for the PID, BDT1 and BDT2 efficiencies, respectively.

\section{Systematic uncertainties}
\label{sec:systs}
The most prominent systematic uncertainties arise from the limited knowledge of the material in the tracking system, the corrections to the PID, BDT1 and BDT2 efficiencies, and the fit model. Note that any possible systematic uncertainties arising from other sources, such as trigger efficiencies, are assumed to be similar enough between the signal and normalisation modes and effectively cancel out in the ratio of efficiencies in Eq.~\ref{eq:BFeq}.

The signal mode has four muons, whilst the normalisation mode has two muons and two pions which can undergo hadronic interactions in the detector material. The $10\%$ uncertainty in the knowledge of the detector material~\cite{LHCb-DP-2013-002} corresponds to a $1.4\%$ uncertainty in the pion detection efficiency and hence a $2.8\%$ uncertainty on $\mathcal{R}$. 
 
The performance of the PID requirement for the normalisation is checked and corrected using the \decay{\psitwos}{\jpsi\pip\pim} peak in data. The statistical uncertainty of the data efficiency is propagated to the systematic uncertainty, corresponding to 0.04\% and 0.03\% for Run~1 and Run~2, respectively. 
The BDT1 efficiency ratio between data and simulation is corrected using the information from the \mbox{\decay{\chicone(1P)}{\jpsi\mup\mun}} decay observed in data (Fig.~\ref{fig:jpsimumuSpectrum}). The statistical uncertainty of the efficiency ratio of the BDT1 selection between data and simulation is taken as a systematic uncertainty, corresponding to 2.80\% and 0.80\% for Run~1 and Run~2, respectively.
For the BDT2 selection, the ratio of the efficiency between the signal and normalisation simulation samples is used to correct the total efficiency. Half of the deviation of the data to simulation ratio from unity is taken as a systematic uncertainty, corresponding to 1.04\% and 0.92\% for Run~1 and Run~2, respectively.

To estimate the uncertainties related to the modelling of both the signal and background, the probability density functions used are varied and the yields redetermined in each case. 
For the baseline fit to the normalisation mode, the background is modelled with a first-order polynomial function, and, as alternatives, with a second-order polynomial function or an exponential function. 
The largest change in normalisation yield is taken as a systematic uncertainty, corresponding to 0.99\% and 4.55\% for Run~1 and Run~2, respectively. 
A similar procedure is carried out for the background model in the signal sample, where the baseline fit function is a first-order polynomial, and alternative functions considered are a second-order polynomial function and an exponential function. The biggest change in yield results in a systematic uncertainty of 0.64\% and 0.18\% for Run~1 and Run~2 respectively.

The effect of varying the signal model is also quantified. The change in signal yield when using the convolution of a BW function with a Crystal Ball function instead of the baseline DSCB is investigated, and results in a systematic uncertainty of 1.55\% and 2.68\% for Run~1 and Run~2 respectively. 

The systematic uncertainties are classified according to whether they are correlated between data-taking periods. The correlated uncertainties include those related to hadronic interactions and the agreement of PID performance between data and simulation. The uncorrelated uncertainties comprise those associated with the two BDT selections and with signal and background modelling. A summary of the systematic uncertainties is given in Table~\ref{tab:systSummary}, together with the total systematic uncertainty per data-taking period, separated into uncorrelated and correlated components. 
\begin{table}[tb]
    \centering
    \caption{Sources of relative systematic uncertainty on $\mathcal{R}$ for each data-taking period, separated into correlated and uncorrelated components.}
    \label{tab:systSummary}
    \begin{tabular}{ccc}
        \multirow{2}{*}{Source} & \multicolumn{2}{c}{Uncertainty (\%)} \\
        & Run~1 & Run~2 \\
        \hline
         Hadronic interactions                  & \multicolumn{2}{c}{2.80} \\
         PID agreement data/simulation              & 0.04 & 0.03 \\
         \hline
         Total correlated  & \multicolumn{2}{c}{2.80}\\
         \hline\hline
         BDT1                                & 2.80 & 0.80 \\
         $\jpsi \pip\pim$ background model & 0.99 & 4.55 \\
         $\jpsi \mup \mun$ background model        & 0.64 & 0.18 \\
        $\jpsi \mup \mun$  signal model            & 1.55 & 2.68 \\
         \hline
         Total uncorrelated  & 3.41 & 5.34\\
    \end{tabular}
\end{table}

\section{Results}
Using the total efficiency, Eq.~\ref{eq:BFeq} can be evaluated for both data-taking periods. The results are
\begin{align*}
    \mathcal{R}_{\rm Run\,1} &= \left(1.54\pm0.68\pm0.05\pm0.04\right)\times10^{-3}, \\
    \mathcal{R}_{\rm Run\,2} &= \left(1.72\pm0.35\pm0.09\pm0.05\right)\times10^{-3},
\end{align*}
for Run~1 and Run~2, respectively. In the expressions above, the first uncertainty is statistical, and the second (third) is the component of the systematic uncertainty that is uncorrelated (correlated) between runs. In order to combine the data-taking periods to obtain the final result, a weighted average is taken accounting for correlations using the best linear unbiased estimator technique~\cite{Nisius:2020jmf}.
\section{Conclusions}
The combined result for the ratio of branching fractions, relative to the \decay{\theX}{\jpsi\pip\pim} normalisation mode (defined in Eq.~\ref{eq:BFeq}), is
\begin{align*}
    \mathcal{R} = \left(1.68\pm 0.32\pm 0.05\right)\times10^{-3},
\end{align*}
where the first uncertainty includes both statistical and uncorrelated systematic contributions, and the second represents the correlated systematic contribution. Using $\mathcal{B}(\theX \rightarrow \jpsi\pip\pim) = (4.3 \pm 1.4) \, \%$ \cite{PDG2024} this corresponds to \mbox{$\mathcal{B}(\decay{\theX}{\jpsi\mup\mun}) = \left(7.2 \pm 2.7\right) \times 10^{-5}$}. The value agrees with the prediction $\mathcal{B}(\decay{\theX}{\jpsi\mup\mun}) = \left(4.2 \pm 1.7\right) \times 10^{-5}$ given in Ref.~\cite{Colangelo:2025yud} which considers only the Dalitz contribution to this final state. This first observation may help to elucidate the nature of the \theX state, and identify potential avenues for future investigation. Furthermore, with a larger dataset, it will be possible to study the form factors of the decay and potentially disentangle the contributions from the virtual photon and \rhoz or \Pomega mesons. It would also allow the enhancement consistent with the $\chi_{c0}(3915)$ state to be investigated in more detail, in order to determine which states are present in this mass region.

\section*{Acknowledgements}
%
%
\noindent We express our gratitude to our colleagues in the CERN
accelerator departments for the excellent performance of the LHC. We
thank the technical and administrative staff at the LHCb
institutes.
We acknowledge support from CERN and from the national agencies:
ARC (Australia);
CAPES, CNPq, FAPERJ and FINEP (Brazil); 
MOST and NSFC (China); 
CNRS/IN2P3 (France); 
BMFTR, DFG and MPG (Germany);
INFN (Italy); 
NWO (Netherlands); 
MNiSW and NCN (Poland); 
MCID/IFA (Romania); 
MICIU and AEI (Spain);
SNSF and SER (Switzerland); 
NASU (Ukraine); 
STFC (United Kingdom); 
DOE NP and NSF (USA).
We acknowledge the computing resources that are provided by ARDC (Australia), 
CBPF (Brazil),
CERN, 
IHEP and LZU (China),
IN2P3 (France), 
KIT and DESY (Germany), 
INFN (Italy), 
SURF (Netherlands),
Polish WLCG (Poland),
IFIN-HH (Romania), 
PIC (Spain), CSCS (Switzerland), 
and GridPP (United Kingdom).
We are indebted to the communities behind the multiple open-source
software packages on which we depend.
Individual groups or members have received support from
Key Research Program of Frontier Sciences of CAS, CAS PIFI, CAS CCEPP, 
Minciencias (Colombia);
EPLANET, Marie Sk\l{}odowska-Curie Actions, ERC and NextGenerationEU (European Union);
A*MIDEX, ANR, IPhU and Labex P2IO, and R\'{e}gion Auvergne-Rh\^{o}ne-Alpes (France);
Alexander-von-Humboldt Foundation (Germany);
ICSC (Italy); 
Severo Ochoa and Mar\'ia de Maeztu Units of Excellence, GVA, XuntaGal, GENCAT, InTalent-Inditex and Prog.~Atracci\'on Talento CM (Spain);
SRC (Sweden);
the Leverhulme Trust, the Royal Society and UKRI (United Kingdom).

\addcontentsline{toc}{section}{References}
\bibliographystyle{LHCb}
\bibliography{main,standard,LHCb-PAPER,LHCb-CONF,LHCb-DP,LHCb-TDR}

\newpage
\centerline
{\large\bf LHCb collaboration}
\begin
{flushleft}
\small
R.~Aaij$^{38}$\lhcborcid{0000-0003-0533-1952},
A.S.W.~Abdelmotteleb$^{58}$\lhcborcid{0000-0001-7905-0542},
C.~Abellan~Beteta$^{52}$\lhcborcid{0009-0009-0869-6798},
F.~Abudin{\'e}n$^{58}$\lhcborcid{0000-0002-6737-3528},
T.~Ackernley$^{62}$\lhcborcid{0000-0002-5951-3498},
A. A. ~Adefisoye$^{70}$\lhcborcid{0000-0003-2448-1550},
B.~Adeva$^{48}$\lhcborcid{0000-0001-9756-3712},
M.~Adinolfi$^{56}$\lhcborcid{0000-0002-1326-1264},
P.~Adlarson$^{86}$\lhcborcid{0000-0001-6280-3851},
C.~Agapopoulou$^{14}$\lhcborcid{0000-0002-2368-0147}
% ...

{\footnotesize \it

% ...
$^{14}$Universit{\'e} Paris-Saclay, CNRS/IN2P3, IJCLab, Orsay, France\\
% ...
$^{38}$Nikhef National Institute for Subatomic Physics, Amsterdam, Netherlands\\
% ...
$^{48}$Instituto Galego de F{\'\i}sica de Altas Enerx{\'\i}as (IGFAE), Universidade de Santiago de Compostela, Santiago de Compostela, Spain\\
% ...
$^{52}$Physik-Institut, Universit{\"a}t Z{\"u}rich, Z{\"u}rich, Switzerland\\
% ...
$^{56}$H.H. Wills Physics Laboratory, University of Bristol, Bristol, United Kingdom\\
% ...
$^{58}$Department of Physics, University of Warwick, Coventry, United Kingdom\\
% ...
$^{61}$School of Physics and Astronomy, University of Glasgow, Glasgow, United Kingdom\\
$^{62}$Oliver Lodge Laboratory, University of Liverpool, Liverpool, United Kingdom\\
% ...
$^{70}$Syracuse University, Syracuse, NY, United States\\
% ...
$^{86}$Department of Physics and Astronomy, Uppsala University, Uppsala, Sweden, associated to $^{61}$\\
% ...
\medskip
}
\end{flushleft}

\end{document}